\documentclass[11pt]{article}
\usepackage{psfig}
\usepackage{amssymb}

\begin{document}
\begin{flushright}
BI-TH-97/38 \\
September 1997
\end{flushright}

\begin{center}
\vspace{24pt}

{\Large \bf Beyond the $c=1$ Barrier in Two-Dimensional \\
            Quantum Gravity}
\vspace{24pt}

{\large \sl Gudmar Thorleifsson \,{\rm  and}\,
            Bengt Petersson}
\vspace{6pt}

Fakult\"{a}t f\"{u}r Physik, Universit\"{a}t Bielefeld,
 D-33615 Bielefeld, Germany

\vspace{10pt}

\begin{abstract}
We introduce a simple model of {\it touching} random
surfaces, by adding a chemical potential $\rho$ for ``minimal necks'', 
and study this model numerically coupled to a Gaussian model in
$d$--dimensions (for central charge $c = d = 0$, 1 and 2). 
For $c \leq 1$, this model has a phase transition to 
branched polymers, for sufficiently large $\rho$.
For $c = 2$, however, the extensive simulations indicate that
this transition is replaced by a cross-over behavior on finite
lattices --- the model is always in the branched polymer phase.
This supports recent speculations that, in $2d$--gravity, the 
behavior observe in simulations for $c \leq 1$, 
is dominated by finite size effects, which are exponentially 
enhanced as $c\rightarrow 1^{+}$.
\end{abstract}

\end{center}

\vspace{15pt}

\section{Introduction}

\noindent
When conformally invariant matter ${\cal S}_M(X)$
is coupled to two-dimensional quantum gravity:
\begin{equation}
 {\cal Z}(\mu) \;=\; \int {\cal D}g \, {\cal D}X
   \; {\rm e}^{\;\textstyle  - \mu\,\int {\rm d}^2\xi \sqrt{|g|}
   - {\cal S}_m(X;g) } \;,
\end{equation}
this breaks down when the matter {\it central charge} $c$ becomes 
larger than {\it one}.  We get unphysical {\it complex} critical 
exponents, such as the {\it string susceptibility exponent}
$\gamma_s$: ${\cal Z}(\mu)~\sim~(\mu_c-\mu)^{2-\gamma_s}$; 
given by the {\it KPZ}--scaling relation:
\begin{equation}
  \gamma_s \;=\; \frac{1}{12} \left ( c - 1
   - \sqrt{(c-25)(c-1)} \right )
\end{equation}
Hence, predictions of continuum theories become meaningless
for $c>1$.  This puzzle, which is related to the
occurrence of {\it tachyons} in bosonic string theories in $d>2$, 
still remains a challenging problem in $2d$--gravity.

Discretized models of $2d$--gravity are, on the other hand, well 
defined for $c>1$, and suitable for studying this problem.
Simplicial gravity, alias dynamical triangulations, is a
discretization of quantum gravity with integrations over metrics
replaced by all possible gluing's of {\it simplices}
into piecewise linear manifolds $T$:
\begin{equation}
 {\cal Z} \;=\; \sum_A {\rm e}^{\;-\mu A}
   \; \sum_{T\in {\cal T}(A)} {\cal Z}_M \;.
\end{equation}
$A$ is the area of the surface, ${\cal Z}_M$ the (discretized) matter 
partition function; for example, a $d$--dimensional Gaussian model 
(bosonic string theory with $c=d$):
\begin{equation}
 {\cal Z}_M \;=\;
   \int {\rm d}^d x \; \delta(x_{cm}) \; {\rm e}^{
   \textstyle  - \sum_{<ij>} (\vec{x}_i - \vec{x}_j)^2} \;,
\end{equation}
and ${\cal T}$ is an appropriate {\it class} of triangulations; 
different classes amount to different discretizations of the 
manifolds, but should yield the same continuum theory.  
Commonly used are {\it combinatorial} (${\cal T}_C$) and
{\it degenerate} (${\cal T}_D$) triangulations.

Models of dynamical triangulations have been studied
extensively, both as {\it matrix models} (for $c\leq 1$) 
and using numerical simulations.  What have we learned
so far: 
\begin{itemize}
 \item
  For $c\leq 1$ the models are well understood; $\gamma_s$ agrees 
  with the {\it KPZ}--scaling and the (internal)
  fractal dimension of the 
  triangulations ($A(r) \sim r^{d_H}$) 
  is $d_H \approx 4$ (still somewhat controversial).
 \item
  For $c \gtrsim 5$ the dominant triangulations are 
  {\it branched polymers} (bubbles glued together in a tree-like
  structure) with $\gamma_s~=~1/2$ and $d_H~=~2$.
 \item
  But, for $1 < c \; \lesssim \;5$ the situation is still unclear.  
  Numerical simulations indicate a {\it smooth} cross-over to 
  the branched polymer phase as $c$ increases.
 \end{itemize}
Is this due to very big finite-size effects \cite{ref2}, 
or is there a different critical behavior for $1 < c \; \lesssim \;5$ ?

\section{Touching random surfaces}

\noindent
A conjecture for the observed $c>1$ behavior, was put
forward in \cite{ref3}:
 ``For $c>1$ the dynamical triangulation model is {\it always}
 in a branched polymer phase.  But finite size effects are
 {\it exponentially enhanced} as $c \rightarrow 1^+$,
 due to the influence of the $c=1$ fixed point
 (which becomes {\it complex} for $c>1$).''

This is based on a large--$N$ renormalization group analysis
of a matrix model including ``touching'' interactions:
\begin{equation}
 {\cal Z} \;=\; \int {\rm d}M
  \; {\rm e}^{\;\textstyle  -N \, {\rm tr} ( M^2 + g M^4 ) -
  x\,( {\rm tr} (M^2)  )^2 } \;.
\end{equation}
For $c\leq 1$ this model has
a transition to branched polymers at a critical
value of the {\it touching} coupling $x$ \cite{ref4}.
For $c>1$, however, this fixed point moves into the
complex plane; but it still influences the {\it RG}--flow's when
$c$ is not too big.

How do we verify this conjecture?
We introduce a simple model of
touching random surfaces, adding a {\it chemical potential}
$\rho$ for {\it minimal necks} $n_m$ on the surface.
As we work with degenerate triangulations, a minimal neck is a
vertex connected to itself {\it via} a link (a {\it tadpole}
in the dual graph).
The (fixed area) partition function is:
\begin{equation}
 {\cal Z}_A(\rho) \;=\;
  \sum_{T \in {\cal T}_D} {\rm e}^{\textstyle \; \rho n_m}
   \; {\cal Z}_M \;.
\end{equation}

We have simulated this model for $c \leq 2$, using 0, 1 and
2 Gaussian models, on surfaces up to 8000 triangles.
Our goal is to verify the existence of a transition to branched
polymers for $c\leq 1$, and to see if this transition
still exists for $c>1$.  Or, alternatively, is it replaced by
{\it cross-over} behavior on finite lattices.

To study the phase structure of we measure the
second derivative of the free energy:
$C_A \;=\; A^{-1}\;\partial^2 \log{{\cal Z}_A}/ \partial \rho^2$,
and the string susceptibility exponent $\gamma_s$.
The latter is obtained from the distribution of {\it baby universes}
on the surface, using the large--$A$ behavior of the partition function:
${\cal Z}_A \;\approx\; {\rm e}^{\mu_c A}  A^{\gamma_s - 3}$.
For $c=1$ this behavior is modified by logarithmic
corrections,
${\cal Z}_A \;\approx\; {\rm e}^{\mu_c A}  A^{\gamma_s - 3}
\log^{\alpha}A$,
 --- including them is essential to extract the
correct $\gamma_s$ numerically \cite{ref5}.
\begin{figure}
\begin{center} \mbox{
\psfig{figure=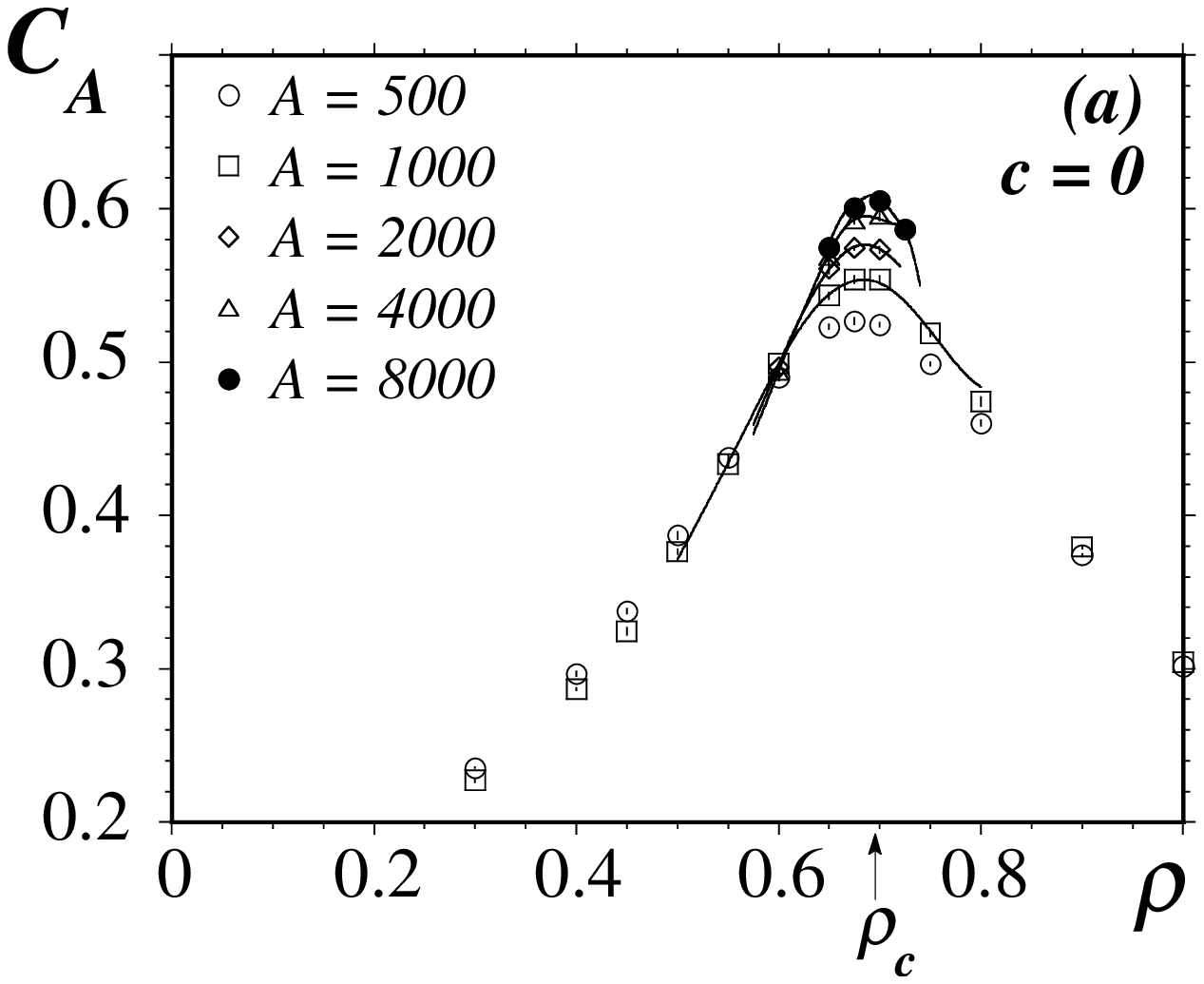,width=2.4in} \hfill \hspace{0.2in}
\psfig{figure=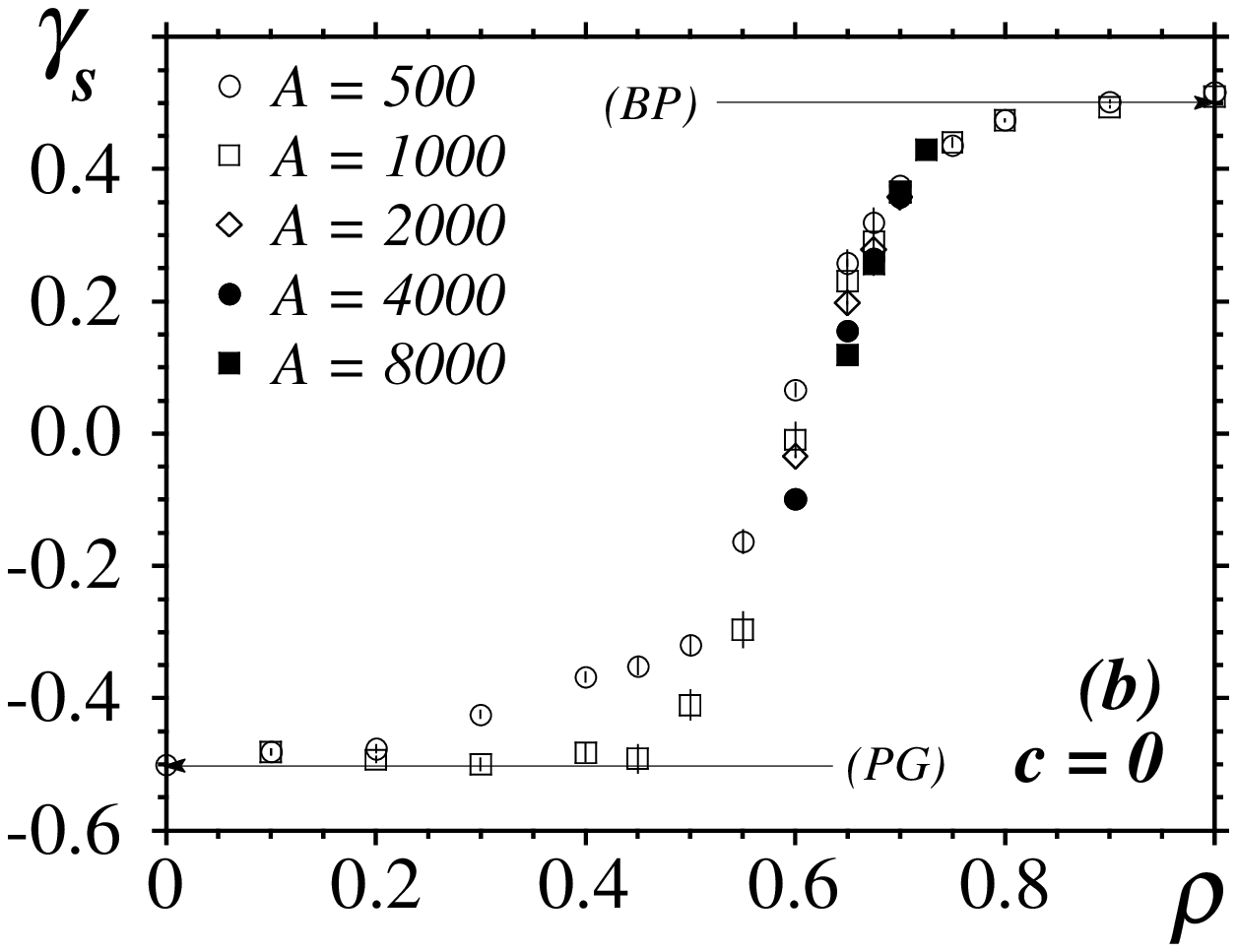,width=2.4in}
} \end{center}
\label{fig1}
\caption{$C_A$ and $\gamma_s$ for $c=0$.}
\end{figure}
\begin{figure}
\begin{center} \mbox{
\psfig{figure=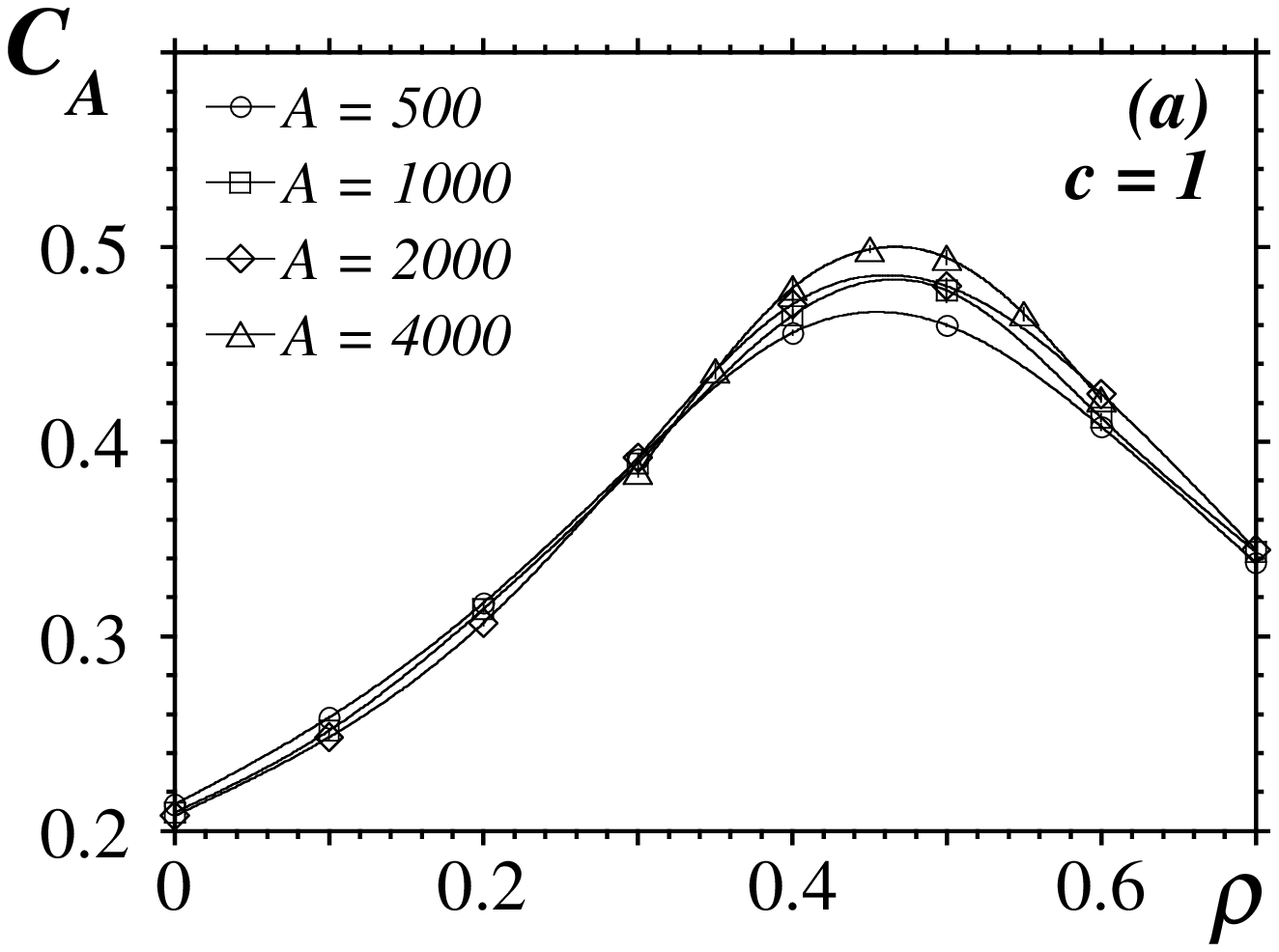,width=2.4in} \hfill \hspace{0.2in}
\psfig{figure=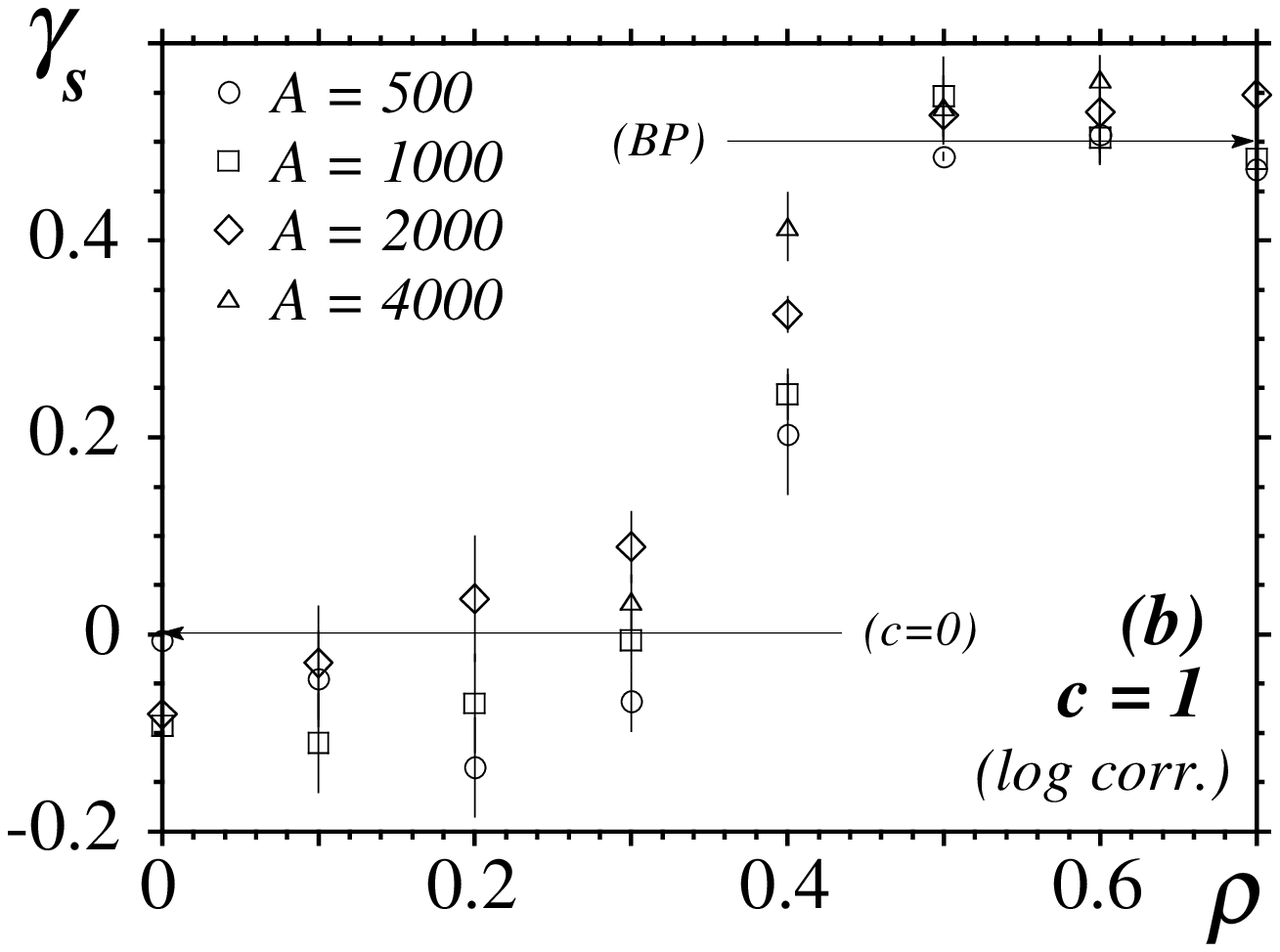,width=2.4in}
} \end{center}
\label{fig2}
\caption{$C_A$ and $\gamma_s$ for $c=1$.}
\end{figure}
\begin{figure}
\begin{center} \mbox{
\psfig{figure=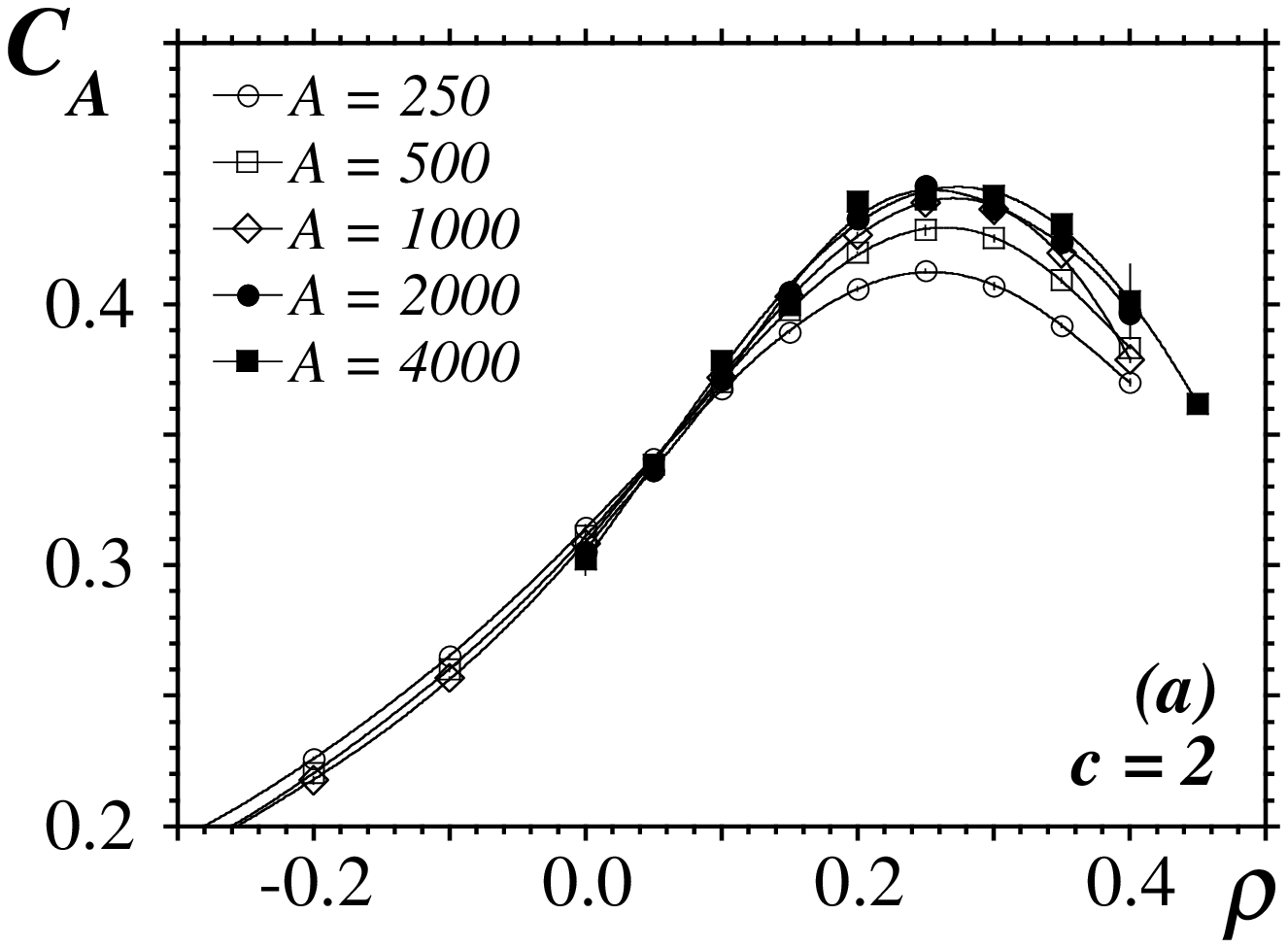,width=2.4in} \hfill \hspace{0.2in}
\psfig{figure=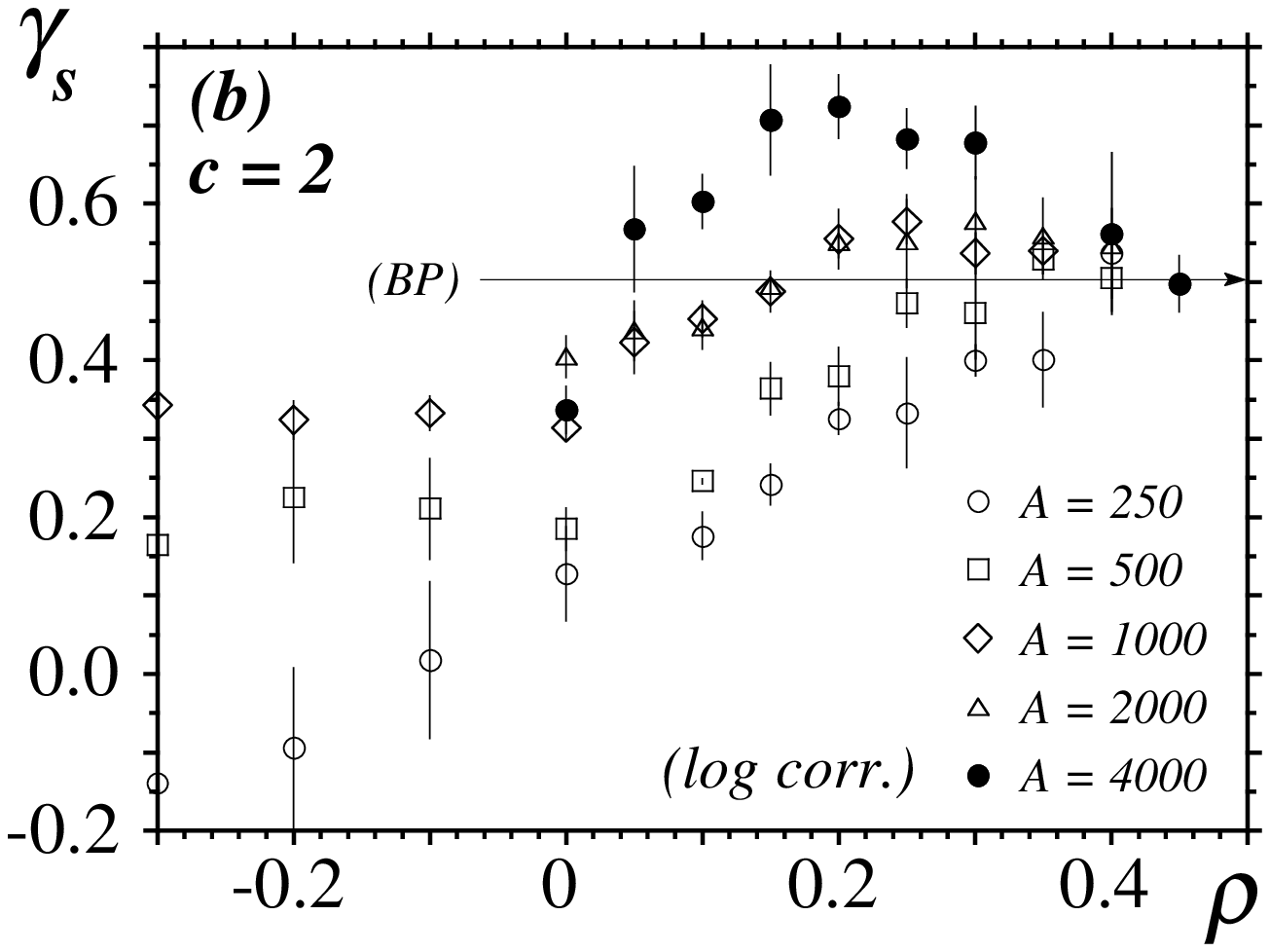,width=2.4in}
} \end{center}
\label{fig3}
\caption{$C_A$ and $\gamma_s$ for $c=2$.}
\end{figure}

\section{Results}

\noindent
For $c=0$ (pure gravity) we see a clear signal of a
phase transition.  There is a peak in $C_A$,
which gets sharper as $A$ increases, but does not diverge
(Fig.~1{\it a}).  Finite size scaling of the peak
($C_A \sim c_0 + c_1 A^{\alpha / \nu d_H}$) gives:
$\rho_c = 0.695(5)$ and $\alpha = -1.07(11)$, assuming
hyper-scaling is valid ($\alpha = 2-\nu d_H$).
(Note that this $\nu$ is related to the touching
interaction; hence $\nu \neq 1/d_H$).
At the same value of $\rho_c$ there is a sharp transition in
$\gamma_s$ from its pure gravity value, $\gamma_s(PG)= -1/2$,
to branched polymers, $\gamma_s(BP) = 1/2$ (Fig.~1{\it b}).

We observe a similar behavior for $c = 1$ (Figs. 2{\it a} and {\it b}):
a non-divergent peak in $C_A$, with
$\rho_c = 0.45(1)$ and $\alpha = -0.8(2)$,
accompanied by a transition to branched 
polymers  in $\gamma_s$.  
In this case, $\gamma_s$ is extracted using logarithmic corrections,
with $\alpha$ as a free parameter.  Below $\rho_c$, $\alpha \approx
-1$, whereas $\alpha \approx 0$ for branched polymers.

For $c>1$, on the other hand, the behavior is different. 
We still observe a peak in $C_A$ (Fig.~3{\it a}), 
but it saturates faster than for $c \leq 1$. 
In fact, $\alpha/\nu d_H < -1$, which implies, if this is
a phase transition, that hyper-scaling is violated.
And, more important, there is {\it no} indication of a phase transition in 
$\gamma_s$, only a smooth cross-over to branched polymers, 
which seems to disappear as $A \rightarrow \infty$ 
(Fig.~3{\it b}).  This is independent of the corrections 
included in extracting $\gamma_s$.  This behavior is,
in our opinion, not compatible with the existence of a phase transition, 
and we conclude that there is only a branched polymer phase.
This strongly supports the conjecture in \cite{ref3}
about the nature of the $c=1$ ``barrier''.

\end{document}